\documentclass[graybox]{svmult}
\usepackage{epsfig,color}
\usepackage{mathptmx}       
\usepackage{helvet}         
\usepackage{courier}        
\usepackage{type1cm}        
\usepackage{makeidx}         
\usepackage{graphicx}        
\usepackage{multicol}        
\usepackage[bottom]{footmisc}
\makeindex             
\usepackage{natbib}
\begin{document}
\bibliographystyle{spbasic}

\title{Stellar Tidal Streams in External Galaxies}
\author{Jeffrey L. Carlin, Rachael L. Beaton, David Mart\'\i nez-Delgado, and R. Jay Gabany}
\institute{Jeffrey L. Carlin: Rensselaer Polytechnic Institute, Troy, New York, USA, and Earlham College, Richmond, Indiana, USA; \email{jeffreylcarlin@gmail.com}\\
Rachael L. Beaton: The Observatories of the Carnegie Institution for Science, Pasadena, CA, USA; \email{rachael.l.beaton@gmail.com}\\
David Mart\'\i nez-Delgado: Astronomisches Rechen-Institut (ARI) am Zentrum fuer Astronomie der Universitaet Heidelberg (ZAH), Germany; \email{delgado@ari.uni-heidelberg.de}\\
R. Jay Gabany: Black Bird Observatory II, Alder Springs, California, USA; \email{rj2010@cosmotography.com}}
%
%
\maketitle

\abstract{ In order to place the highly substructured stellar halos 
  of the Milky Way and M31 in a larger context of hierarchical galaxy formation, 
  it is necessary to understand the prevalence and properties of tidal substructure around external galaxies. 
 This chapter details the current state of our observational knowledge of streams in galaxies in and beyond the Local Group,
  which are studied both in resolved stellar populations and in integrated light. 
 Modeling of individual streams in extragalactic systems is hampered by our inability 
  to obtain resolved stellar kinematics in the streams,
  though many streams contain alternate luminous kinematic tracers, such as globular clusters or planetary nebulae.
 We compare the observed structures to the predictions of models of galactic halo formation,
  which provide insight in the number and properties of streams expected around Milky Way like galaxies.
 More specifically, we discuss the inferences that can be made about stream progenitors based only on observed morphologies.
 We expand our discussion to consider hierarchical accretion at lower mass scales, 
  in particular the observational evidence that substructure exists on smaller mass scales
  and the effects accretion events may have on the evolution of dwarf galaxies (satellite or isolated).
 Lastly, we discuss potential correlations between the presence of substructure in the halo
  and the structural properties of the disk.
 While many exciting discoveries have been made of tidal substructures around external galaxies, 
  the ``global'' questions of galaxy formation and evolution via hierarchical accretion 
  await a more complete census of the low surface brightness outskirts of galaxies in and beyond the Local Group. }

\section{Introduction}

Low surface brightness\index{surface brightness} features, including ``tails" and ``bridges," are visible in highly disturbed galaxies that result from a major merger or strong encounter \citep{arp1966, toomre1972}. 
Similarly, diffuse stellar streams and shells\index{shells} around massive elliptical galaxies have been known for decades, 
 and are attributed either to the accretion of smaller disk galaxies \citep{quinn1984} 
 or to recent, ``in-situ''  star formation from gas that was already contained within the galaxy \citep{fabian1980}. 
\citet{schweizer1988} extended these observations to early spiral galaxies, and suggested that the ``ripples," as they called the shell-like features, were formed through mass transfer from nearby galaxies, in addition to wholesale mergers.

It has only been in recent years, with the advent of wide-area, deep photometric surveys, 
 that the number and variety of stellar substructures (resulting from the tidal disruption of dwarf galaxies\index{dwarf galaxies} and globular clusters\index{globular clusters} in ``minor mergers'') threading the halos of the Milky Way and Andromeda (M31)\index{Andromeda (M31) galaxy} galaxies has become apparent. 
These stellar substructures can be studied in detail and together describe the hierarchical\index{hierarchical galaxy formation} merging history 
 of the two dominant galaxies in the Local Group. 
However, placing them in the broader context of cosmological galaxy formation models\index{cosmological simulations} requires a 
 more general picture of halo substructure only feasible via the exploration of a large number of more distant systems.
Only with such a dataset in hand is it possible to determine whether the Milky Way and M\,31
 have experienced `typical' or `atypical' merging histories \citep[e.g.,][]{mutch2011}. 

Models \citep[e.g.,][]{johnston08, cooper2010} predict that a survey reaching a surface brightness\index{surface brightness} of $\sim$29~mag~arcsec$^{-2}$ 
 around $\sim$100 galaxies outside the Local Group should reveal many tidal features, perhaps as much as one detectable stream per galaxy. 
However, a suitably deep data set that is sensitive to low surface brightness features in a large number of galaxies does not yet exist, 
 leaving the observational evidence needed to test these predictions incomplete. 
In the sections that follow, we will discuss the isolated discoveries of tidal debris structures in external galaxies and their overall utility for elucidating general stellar tidal features.

\section{Stellar streams: detection methods and examples}

Local Group galaxies, including the Milky Way and M31\index{Andromeda (M31) galaxy}, can be dissected star by star, and thus provide a laboratory for understanding the details of hierarchical galaxy formation\index{hierarchical galaxy formation}. However, it is unclear whether the Local Group galaxies represent typical evolution histories. To understand the place of the Milky Way and M31 in the larger context requires studying tidal substructures in galaxies beyond the Local Group, where resolving individual stars is difficult. In this section, we detail some of the methods used to discover tidal debris features, including resolved and unresolved stellar populations, and discuss the unique insights gleaned from some examples of these structures. 

\subsection{Resolved stellar structures in the Local Group}

The era of massive, deep photometric surveys has enabled the detection of tidal debris structures within the Milky Way 
 as stellar overdensities of carefully selected tracers. 
The famous ``Field of Streams''\index{SDSS Field of Streams} image from SDSS\index{Sloan Digital Sky Survey (SDSS)} (\citealt{belokurov2006}; reproduced in Chapter~1) mapped substructure using main sequence turnoff (MSTO) starcounts to surface brightness\index{surface brightness} limits of fainter than $\sim32$~mag~arcsec$^{-2}$. 
The identification of individual members of stellar streams enables extremely low surface brightness features to be detected. 
This can be achieved by kinematical selection of stream members (for example, via spectroscopic velocities)\index{radial velocities}, 
 allowing features in the Galactic halo containing fewer than $\sim1$~red giant branch (RGB)\index{red giant branch (RGB) stars} star per square degree to be identified. 

Likewise, in M\,31\index{Andromeda (M31) galaxy}, photometric selection of metal-poor RGB candidates removes much of the contaminating background, 
 and has allowed detections of features as faint as $\sim30$~mag~arcsec$^{-2}$, 
 including the ``giant stream''\index{Giant Stellar Stream (GSS)} \citep{ibata2001} and other low surface brightness\index{surface brightness} features (e.g., \citealt{ferguson2002}; see Chapter 8 for more about M31). 
When individual stars can be resolved and spectroscopically vetted in M\,31, the measurement of surface brightnesses 
 as faint as $\sim32$ mag~arcsec$^{-2}$ is possible (see, e.g., \citealt{gilbert2012} and other results from the SPLASH survey\index{SPLASH survey}).
Thus, we are able to probe both the Milky Way and M\,31 to depths of $\sim32$~mag~arcsec$^{-2}$, 
 which is deep enough to sample most of the simulated halo substructures seen in simulations (discussed further in Section~\ref{gx_formation_section}).


\begin{figure}[t]
  \begin{center}
   \includegraphics[height=1.0\textwidth, angle=270]{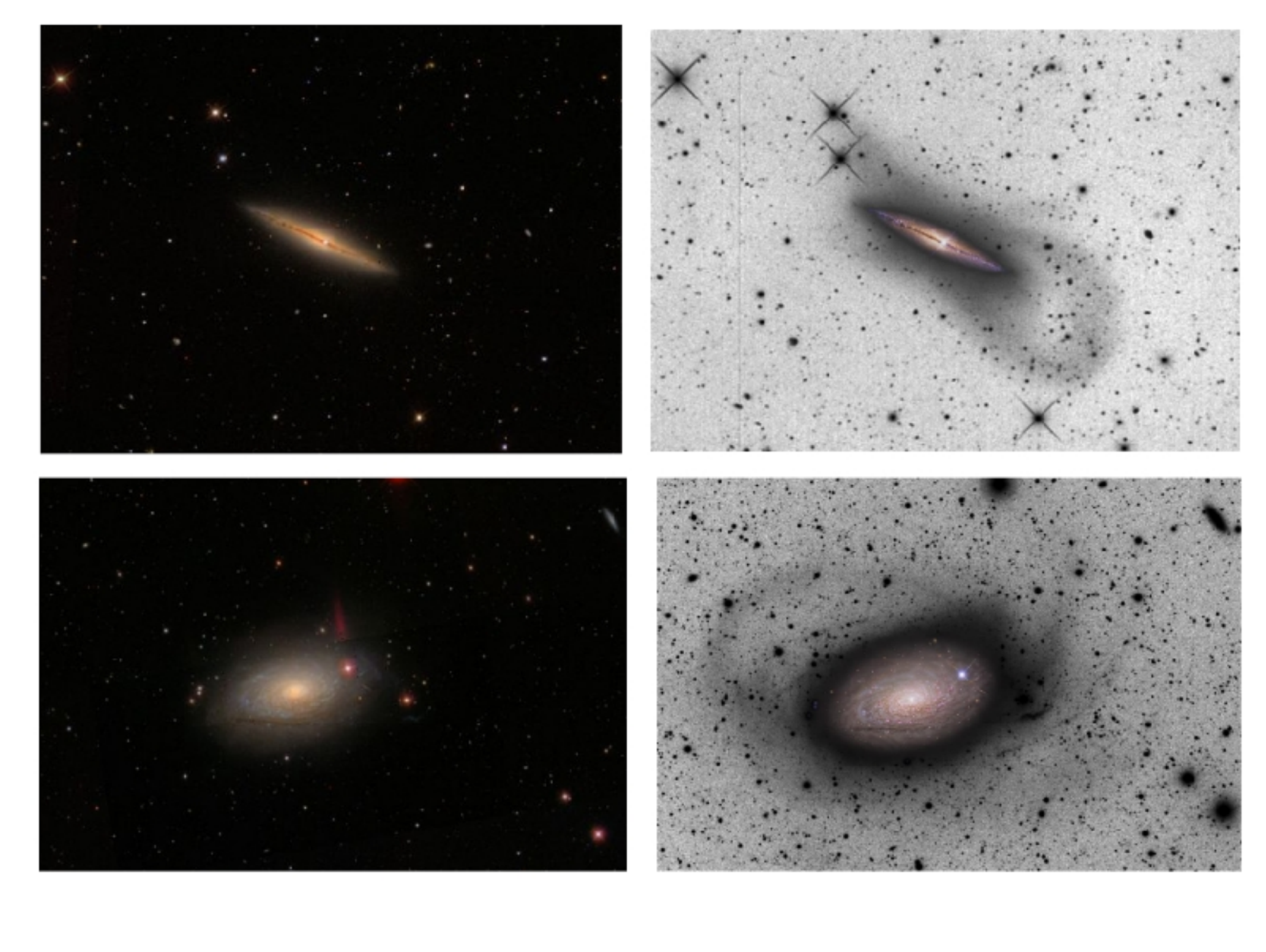}
    \caption{Left panels: images from the Sloan Digital Sky Survey of the nearby galaxies NGC~4013\index{NGC 4013} (top) and M~63\index{M 63} (bottom). 
 These images do not show any obvious signs of tidal streams in the halos of these galaxies at the surface brightness\index{surface brightness} limit of SDSS\index{Sloan Digital Sky Survey (SDSS)}. 
 Deep images of these same galaxies reveal a low-latitude stellar stream around NGC~4013 (upper right panel; \citealt{MD2009}) 
  and a giant tidal stream around the spiral galaxy M~63 (lower-right; \citealt{chonis11}).
 For reference, a color inset of each galaxy's central regions has been inserted atop the deeper images. 
 These images illustrate the value of deep, sensitive imaging (i.e., beyond that of SDSS) for detecting the faint debris structures predicted by theoretical models in external galaxies.}
    \label{n4013_m63}
  \end{center}
\end{figure}

\subsection{Detection methods in and beyond the Local Group}

Current ground-based telescopes are unable to resolve stellar tidal streams around most galaxies beyond the Local Group into individual stars.
Most of the streams that have been found in external galaxies thus appear as elongated, 
 diffuse-light regions extending over several arcminutes on the sky. 
To map such tidal streams requires deep imaging that is also sensitive to extremely faint surface brightness features; 
 the typical surface brightness\index{surface brightness} of known stellar tidal streams is 27~mag~arcsec$^{-2}$ or fainter, 
 depending both on the luminosity of the progenitor\index{progenitors of streams} and the time they were accreted \citep{johnston08}.

Faint tidal features can be identified on sky-limited archival photographic plates using a process called photographic amplification; 
 the surface brightness\index{surface brightness} limit is even fainter if photographically amplified derivatives of several plates are combined together \citep{1981AASPB..27....4M}.  
Using these techniques (photographically or digitally) it is possible to detect extended features to 28~mag~arcsec$^{-2}$ 
 from existing photographic surveys \citep{malin_hadley1997}.
This depth is comparable to that achievable with SDSS\index{Sloan Digital Sky Survey (SDSS)} in Figure~\ref{n4013_m63}.

Small, fast (i.e., low focal-ratio) telescopes (e.g., \citealt{MD2010, vandokkum2014}) and modern CCD cameras are 
 capable of imaging unresolved structures in external galaxies to $\Sigma_{R}\sim$29 mag~arcsec$^{-2}$.
Detecting these faint features requires very dark sky conditions and images taken with exquisite flat-field quality 
 over a relatively large angular scale.
More specifically, stellar streams are typical found at large galactocentric distances (15 kpc~$ < R <$~100 kpc, or farther) 
 and could be found out to a significant portion of the virial radius of the parent galaxy (for the Milky Way or M\,31\index{Andromeda (M31) galaxy}, $R_{\rm virial} \sim 300$ kpc).
Thus, surveys for stellar debris must produce images over large angular scales (from $> 30'$ for systems $< 10$~Mpc to $\sim$15$'$ for systems $\sim$50~Mpc away).  
As an example, the survey strategy employed by \citet{MD2010} uses stacks of multiple deep exposures of each target 
 taken with high throughput clear ``luminance'' filters (transmitting 4000~\AA~$<~\lambda~<$~7000~\AA\, with a near-IR cut-off) 
 with typical exposure times of 7-8 hours.

\begin{figure}[!t]
  \begin{center}
   \includegraphics[height=1.0\textwidth, angle=270]{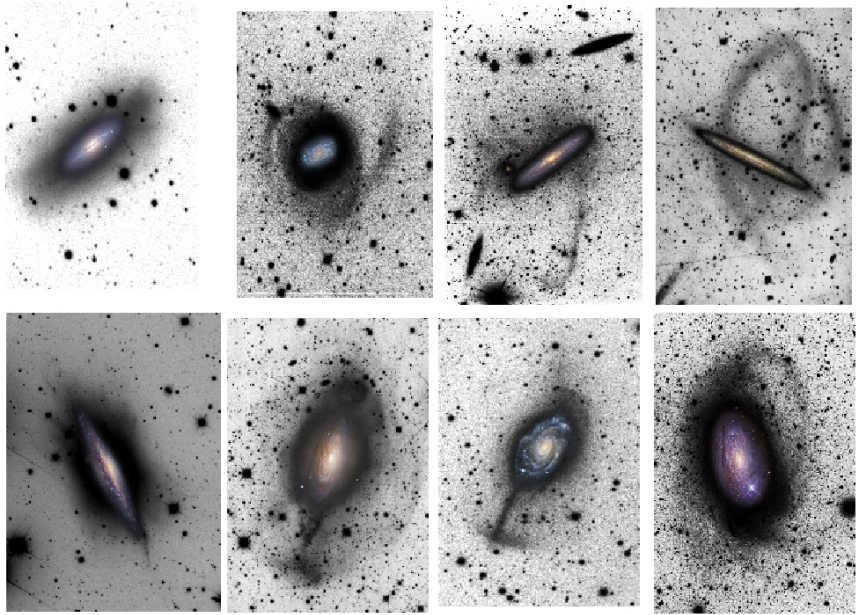}
    \caption{ \footnotesize{Luminance filter images of nearby galaxies from the pilot survey of \citet{MD2008, MD2010} showing large, diffuse light structures in their outskirts. These include tidal streams similar to Sagittarius\index{Sagittarius (Sgr) stream} (upper right panel), giant plumes (middle panels in the top row), partially disrupted satellites (top row, third panel from left), umbrella-shaped\index{umbrellas} tidal debris structures (middle two panels in the bottom row), enormous stellar clouds\index{clouds}, prominent spikes, and large scale, complex inner halos sprinkled with several debris features. A color inset of the disk of each galaxy has been overplotted for reference.  An illustrative comparison of these features to the surviving structures visible in cosmological simulations\index{cosmological simulations} is given in Martinez-Delgado et al. 2010, their Fig 2.}}
    \label{zoo}
  \end{center}
\end{figure}

Recent deep, wide-field imaging surveys have focused on nearby spiral galaxies that were suspected 
 (based on existing data from, e.g., surveys such as POSS-II or SDSS\index{Sloan Digital Sky Survey (SDSS)}, or from amateur astronomical imaging) to contain diffuse-light over-densities.
To date, the combined observational efforts have revealed more than 50 previously undetected 
 stellar structures in galaxies as distant as 80~Mpc.
Figure~\ref{zoo} shows eight such galaxies from the survey of \citet{MD2010}, 
 illustrating the variety of tidal debris features -- both in their morphologies and in their projected radii.
The central disks of the galaxies shown in Figure~\ref{zoo} are of similar overall physical size, 
 but the debris features can span large galactocentric distances.
The morphologies\index{morphologies of debris structures} of the features include great circle\index{great circle streams} streams resembling the Sagittarius stream\index{Sagittarius (Sgr) stream} around our 
 Galaxy (upper right panel of Figure~\ref{zoo}; Sagittarius is discussed in Chapter~2 of this volume), 
 isolated shells\index{shells}, giant clouds\index{clouds} of debris floating within galactic halos, jet-like features emerging from galactic disks, 
 and large-scale diffuse structures that may be related to the remnants of ancient, already thoroughly disrupted satellites. 
The diversity in observed substructure morphology parallels that seen in simulations \citep[e.g.,][]{johnston08, cooper2010}.
In addition to the remains of satellites that are likely completely destroyed, 
 there are a few examples (e.g., \citealt{MD2012, MD2014, amorisco2015}) of surviving satellites caught in the act of tidal disruption, 
 displaying long tails departing from the progenitor satellite\index{progenitors of streams} (i.e., similar in spirit to observations in the Local Group). 
The extraordinary variety of morphological specimens provides strong evidence to 
 support the hierarchical galaxy formation\index{hierarchical galaxy formation} scenarios predicted by cosmological models\index{cosmological simulations} \citep[e.g.,][]{cooper2010}. 

\subsection{Unresolved features beyond the Local Group}

Prior to recent dedicated CCD searches, only a few cases of extragalactic stellar tidal streams have been reported in nearby spiral galaxies.
\citet{malin_hadley1997} found two possible tidal streams surrounding the galaxies M\,83 and M\,104 by using special contrast enhancement techniques on 
 plates obtained for wide-field photographic surveys. 
Shortly thereafter, a study of the nearby, edge-on galaxy NGC~5907\index{NGC 5907} by \citet{shang98} 
 employed deep CCD images to reveal an elliptical loop in the halo of this galaxy, 
 which they believed to be the remains of a tidally disrupted galaxy similar in size to the Sagittarius dwarf galaxy\index{Sagittarius (Sgr) dwarf spheroidal (dSph)} 
 (Sagittarius was at that time a recent discovery in the Milky Way halo). 
Shang et al. also identified another Sagittarius-like dwarf galaxy\index{dwarf galaxies} that they suggested might be interacting with the disk of NGC~5907, 
 causing the observed warp in HI. 
Their photometry reached surface brightnesses\index{surface brightness} of 28.6 and 26.9 mag~arcsec$^{-2}$ in $R$ and $I$-bands, respectively.

\begin{figure}[t]
  \begin{center}
   \includegraphics[height=1.0\textwidth, angle=270]{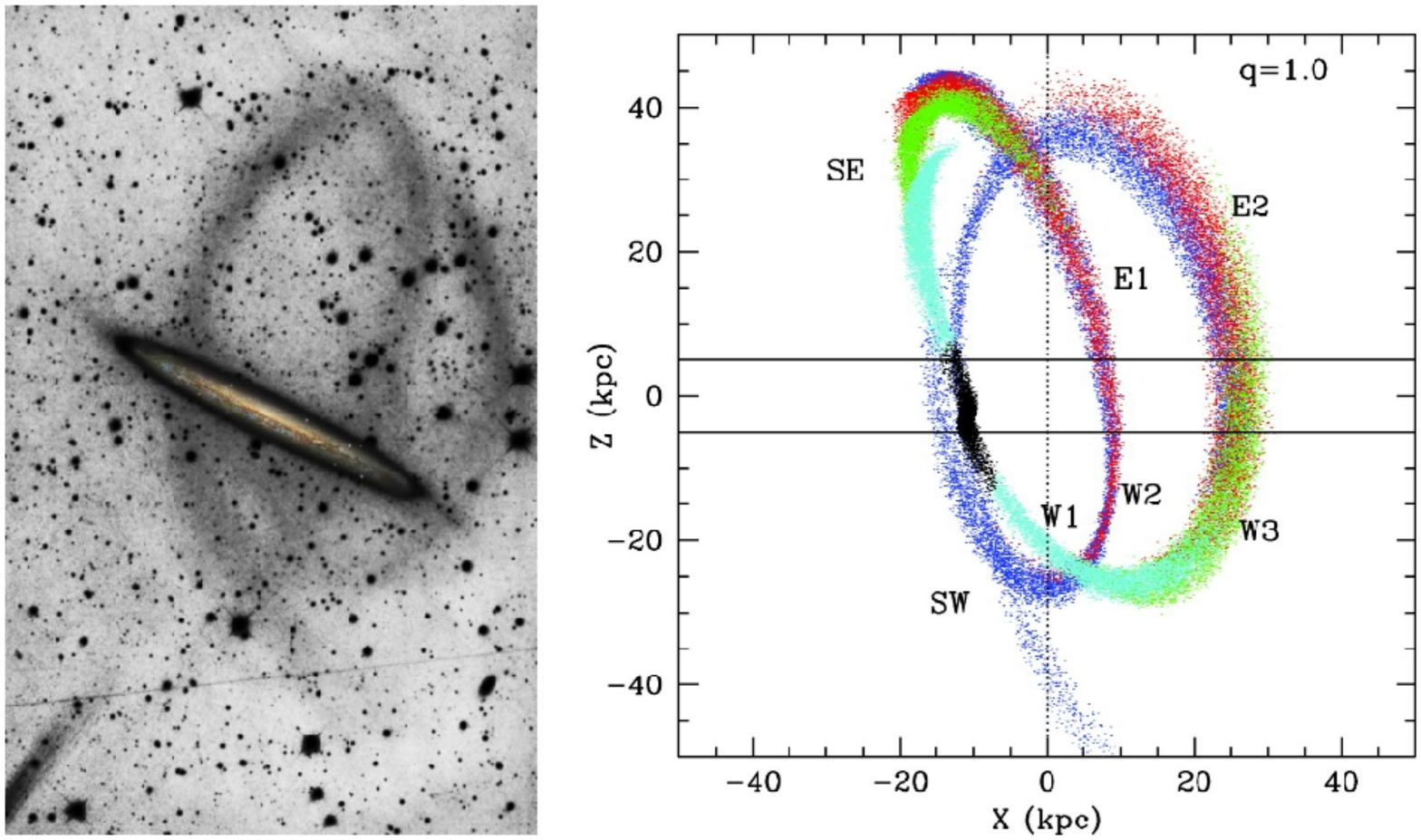}
    \caption{Left: deep image of the stellar tidal stream around NGC~5907\index{NGC 5907} obtained with the 0.5-meter BBO telescope \citep{MD2008}. 
  The great-circle\index{great circle streams} morphology of this system is likely very similar to that of the Sagittarius stream\index{Sagittarius (Sgr) stream} in the Milky Way. 
 Right: $N$-body model\index{N-body simulations} of the NGC~5907 stellar stream. 
 The satellite is realized as a King model with an initial mass, King core and tidal radii of $M=2\times10^8 M_\odot$, $r_c=0.39$ kpc and $r_t=2.7$ kpc, respectively. 
 Different colors denote particles that became unbound after  different pericentric passages, whereas black particles are those that remain bound. 
 The fainter, outer loop material (blue points) became unbound at least 3.6 Gyrs ago. 
 For this particular model the orbital\index{orbit fitting} period is $T_r=0.9$ Gyr.}
    \label{n5907}
  \end{center}
\end{figure}

More recently, NGC\,5907\index{NGC 5907} was imaged by \citet[][Figure~\ref{n5907}]{MD2008}, revealing even fainter features of the stream.
As shown in Figure~\ref{n5907}, this stream 
 is prominently visible as an interwoven, rosette-like structure traversing nearly $720^{\circ}$ around NGC\,5907. 
Detailed $N$-body modeling\index{N-body simulations} suggests that {\it all} of the features can be reproduced from the accretion of a single, low-mass satellite galaxy,
 with the stream tracing two full orbits of its progenitor. 
The presence of such a long stream confirms that a stellar substructure can survive several gigayears, 
 which though predicted by $N$-body simulations\index{N-body simulations} of tidally disrupted stellar systems around the Milky Way \citep[e.g.,][]{law2005,penarrubia2005}
 lacked direct confirmation. 
Interestingly, the $N$-body model of Figure~\ref{n5907} was created using a progenitor\index{progenitors of streams} with orbital parameters similar those found for Sagittarius\index{Sagittarius (Sgr) dwarf spheroidal (dSph)} in the Milky Way (Chapter 2). 
The model suggests that the fainter, outer loop material (blue points in Figure~\ref{n5907}) became unbound from its progenitor at least 3.6~Gyr ago.
The substructure in NGC\,5097\index{NGC 5907} is one of the most striking examples of an external great-circle\index{great circle streams} tidal stream to date.

While some galaxies have {\it great-circle}\index{great circle streams} tidal streams that resemble the Sagittarius stream surrounding 
 our Galaxy (e.g., in NGC 5907: Figure~\ref{n5907}, or NGC~4013 and M63: Figure~\ref{n4013_m63}), 
 others have enormous structures that resemble open umbrellas\index{umbrellas}, and that extend over tens of kiloparsecs (e.g., the middle two panels in the bottom row of Figure~\ref{zoo}). 
These structures are often located on both sides of the host galaxy, and 
 display long narrow shafts that terminate in a giant shell\index{shells} of debris \citep[e.g., NGC~4651;][]{foster2014}\index{NGC 4651}. 
Another umbrella-like feature, dubbed the ``dog leg stream'' \citep{amorisco2015}, has a long narrow spoke (with an embedded progenitor)\index{progenitors of streams} that stretches to a radius of $\sim150$~kpc beyond the center of NGC~1097\index{NGC 1097}, terminating in a ``dog-leg'' that appears like an umbrella\index{umbrellas} feature with one half of the shell missing (note that this system has other narrow plumes visible as well).
With such examples, we are beginning to see real streams around galaxies in the local universe that resemble the menagerie of morphological features\index{morphologies of debris structures} predicted by $\Lambda$-CDM hierarchical structure formation\index{hierarchical galaxy formation} models
\citep[e.g., ][; see also Chapter 6 of this volume]{johnston08, cooper2010}.

While there have been numerous isolated discoveries of debris features around external galaxies, 
 there have been few large-scale systematic surveys to build a comprehensive census of halo substructures. 
Only such a survey can inform simulations by providing estimates of the prevalence of streams of different morphologies, and thus different progenitor masses, orbits, and infall times.\index{progenitors of streams} 
One systematic search by \citet{miskolczi2011} analyzed 474 galaxies in SDSS\index{Sloan Digital Sky Survey (SDSS)} and found clear tidal features around 6\% of the galaxies, 
 with 19\% exhibiting some features above the surface brightness\index{surface brightness} limit of $\sim28$~mag~arcsec$^{-2}$. 
From imaging data in the Canada-France-Hawaii Telescope Legacy Survey, 
 \citet{atkinson13} find tidal features (including both minor and major merger events) around 12\% of the galaxies imaged. 
Given that the $\Lambda$-CDM paradigm predicts that we should see accretion relics around {\it all} Milky Way-sized galaxies, 
 this $\sim10\%$ fraction from the SDSS/CFHT studies would seem to suggest a significant deficit of detected accretion events relative to predictions. 
However, as illustrated in Figure~\ref{n4013_m63} (see also Figure~\ref{johnston_sim_comp}), 
 the surface brightness\index{surface brightness} limits for most large scale surveys are simply too shallow to reveal the complex webs of substructure both predicted in simulations \citep[e.g.,][]{bj2005}
 and observed locally in the Milky Way and M\,31\index{Andromeda (M31) galaxy} (i.e., where $\mu > 28$~mag~arcsec$^{-2}$ can be attained using resolved stellar populations).
However, it is puzzling that no tidal stream currently known in the Milky Way or M31 is remotely as bright as the ``faint limits'' of these large scale searches (Sagittarius\index{Sagittarius (Sgr) stream} -- by far the brightest stream in the Milky Way -- is only $\sim$30 mag~arcsec$^{-2}$ at about 30-40 deg. from the core, according to \citealt{mateo1998sgr}).

\subsection{Resolved structures beyond the Local Group}

\begin{figure}[t]
  \begin{center}
  \includegraphics[width=0.8\textwidth]{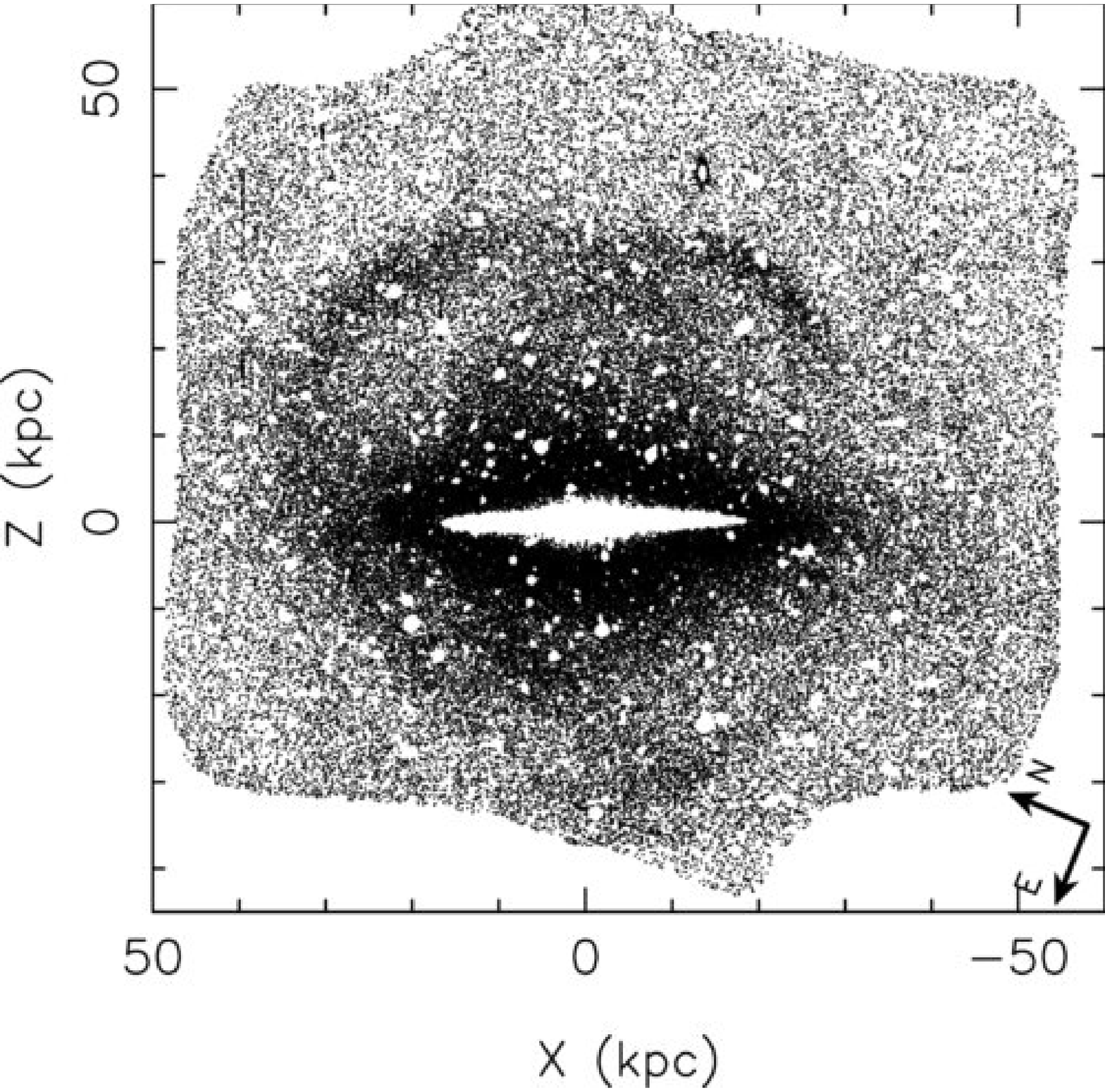}
    \caption{Stellar density map of RGB stars\index{red giant branch (RGB) stars} with magnitudes $25.8\le i_0 \le 27.0$ over a $\sim90\times90$~kpc region surrounding the Milky Way analog galaxy NGC~891\index{NGC 891}. Multiple interlocking loops and arc-like (``great circle'')\index{great circle streams} features are visible over vast regions of the NGC~891 halo. [Figure reproduced by permission from \citet{mouhcine2010}.]}
    \label{n891}
  \end{center}
\end{figure}

With large aperture facilities equipped with wide-field detectors, it is possible to resolve individual stars in some tidal debris structures beyond the Milky Way and M31\index{Andromeda (M31) galaxy}. 
Perhaps the most spectacular example of this from ground-based observations is the Milky Way analog NGC~891\index{NGC 891}, at a distance of $\approx10$~Mpc, 
 which was surveyed by \citet{mouhcine2010} with Suprime-Cam on the 8.2m Subaru telescope. 
With very long ($>11$-hr) exposures, this study resolved stars to $\sim2$ magnitudes below the RGB tip\index{RGB tip stars (TRGB)} in NGC~891, 
 covering a $\sim90\times90$~kpc area around the galaxy down to $i$-band magnitudes fainter than 28th mag (see Figure~\ref{n891}). 
Surface density maps of RGB stars show a complex of features looping throughout the halo of NGC~891\index{NGC 891}, 
 suggesting that the halo of this galaxy contains numerous accretion remnants. The disk of NGC~891 also appears to be ``super-thick'' \citep{mouhcine2010}, providing further evidence of recent accretion. 
Another recent example of deep, ground-based observations is the study by \citet{greggio2014}, who mapped the density of RGB stars\index{red giant branch (RGB) stars} in the halo of the spiral galaxy NGC~253\index{NGC 253} at a distance of 3~Mpc using deep $Z$- and $J$-band imaging from the VISTA telescope.
As a whole, Greggio et al. found that the halo of NGC~253 is fairly homogeneous out to $\sim50$~kpc,  
 with the exception of a $\sim$20~kpc wide shell\index{shells} roughly 28~kpc from the plane that is interpreted to be the result of a recent tidal interaction.
While these ground-based studies are spectacular, the extremely deep, large-area observations required to resolve individual RGB stars in the accretion relics highlight the difficulty (or perhaps impossibility) of doing similar work for large numbers of Milky Way analogs. 

Alternatively, space based telescopes can provide the necessary spatial resolution and depth to trace extra-galactic substructures with individual stars, 
 albeit over significantly smaller angular scales than those on the ground (the HST+ACS\index{Hubble Space Telescope (HST)} field of view is $\sim4'$). 
The HST+ACS GHOSTS survey \citep{radburn_smith2011} resolved stars at the RGB tip\index{RGB tip stars (TRGB)} 
 and used them to map low surface brightness\index{surface brightness} features (to $\Sigma_V \sim 30$~mag~arcsec$^{-2}$) in the outer regions of 14 disk galaxies out to distances of $\sim$17~Mpc.
\citet{bailin2011} began with GHOSTS images from HST for the spiral galaxy NGC~253\index{NGC 253}, and supplemented these with imaging over a much wider field of view with Magellan/IMACS, reaching well below the RGB tip\index{red giant branch (RGB) stars}. 
They estimated the total stellar luminosity of the NGC~253 halo to be roughly twice that of the Milky Way or M31\index{Andromeda (M31) galaxy}, 
 and fit profiles to stellar densities out to $\sim30$~kpc from the galaxy center. 
The shelf-like feature to the south that had been seen by \citet{malin_hadley1997} in photographic plates is clearly visible, 
 as well as other substructure at the $\sim$kpc level. 
Thus, targeted follow-up, ground-based or space-based, for stellar streams discovered in integrated light 
 has the potential to build datasets complementary to those generated en masse for the Milky Way and M\,31\index{Andromeda (M31) galaxy}, 
 reaching appropriate physical spatial scales and surface brightnesses. 

\section{$N$-body modeling of streams}\index{N-body simulations}

We now turn to discussion of what can be learned from the identification and subsequent theoretical modeling of tidal streams in galaxies beyond the Local Group. In many of the examples illustrated in this chapter, we have shown $N$-body models\index{N-body simulations} of disrupting satellites that roughly reproduce the observed morphology of detected streams. For example, for the great-circle\index{great circle streams} tidal stream around NGC~5907\index{NGC 5907} (Figure~\ref{n5907}), an $N$-body model that best replicates the morphology of the observed stream requires a massive Sgr-like galaxy that has spread debris over at least three orbits. If this is the case, then the complex stream structure seen around this galaxy can be entirely explained by a single accretion event. However, due to the difficulty in measuring kinematics for low surface brightness\index{surface brightness} tidal streams (and, indeed, the impossibility of measuring kinematics of individual stars) at several Mpc away, models must be constrained solely by the observed morphology\index{morphologies of debris structures} and the stellar density along the stream. While the panoramic perspective we are afforded of these systems offers many constraints on the properties of the progenitors and their orbits, kinematics will ultimately be needed to fully characterize each accretion event.

It is possible that carefully chosen spectroscopic observations can derive bulk kinematics of some tidal debris features around external galaxies. For example, \citet{sanderson2013} outlined a method to do this for tidal caustics (or ``shells'')\index{shells} via careful choice of spectroscopic fiber positioning and identification of the tell-tale velocity\index{radial velocities} signature. A much more promising avenue is to use intrinsically brighter point-like tracers such as globular clusters\index{globular clusters}, planetary nebulae\index{planetary nebulae (PNe)}, or HII regions\index{HII regions} to elucidate debris structures. The densities of globular clusters have been used \citep{dabrusco2015} to show large structures in the halos of Virgo cluster galaxies that may be evidence of recent accretion events. 

\begin{figure}[t]
  \begin{center}
   \includegraphics[height=1.0\textwidth, angle=270]{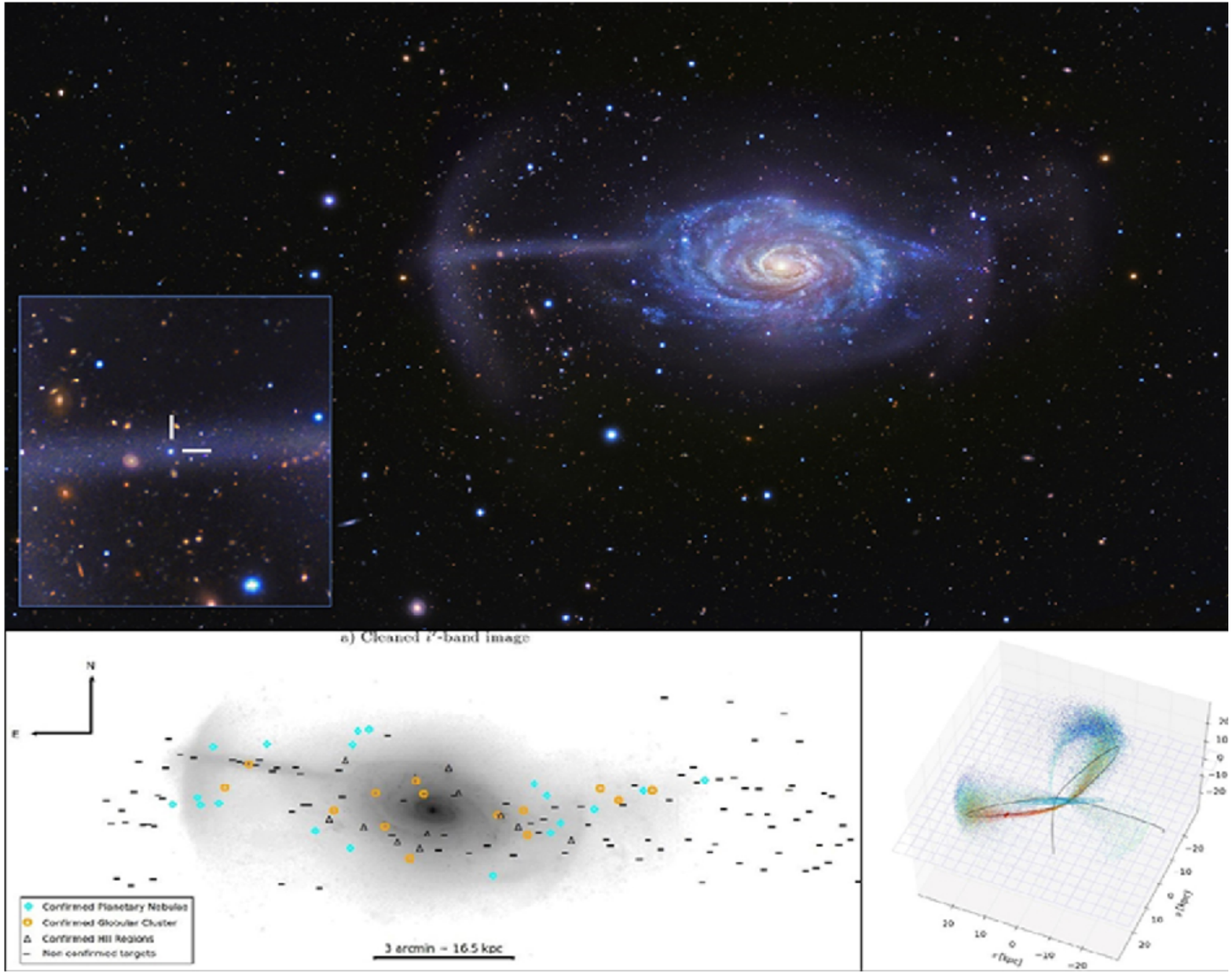}
    \caption{The stellar stream in NGC4651\index{NGC 4651}. Top: color image of the extensive tidal features in NGC4651, including an ``umbrella''\index{umbrellas} and several shells\index{shells} of material. The stream has an exceptionally blue color. Bottom left: several globular clusters\index{globular clusters} (yellow circles), planetary nebulae\index{planetary nebulae (PNe)} (cyan diamonds), and HII regions\index{HII regions} (black triangles) are found along the debris. These bright tracers permit kinematical probes that greatly enhance the ability of $N$-body modeling\index{N-body simulations} (bottom right) to elucidate the physical parameters of the accretion event.
 [Reproduced from \citet{foster2014}.]}
    \label{n4651}
  \end{center}
\end{figure}

An example of a stream in a distant galaxy for which kinematics have been measured and an orbit derived\index{orbit fitting} is that of the Umbrella Galaxy (NGC~4651)\index{NGC 4651}. \citet{foster2014} followed up the low surface-brightness\index{surface brightness} imaging of \citet{MD2010} with even deeper imaging from the Subaru/Suprime-Cam instrument. Figure~\ref{n4651} shows images from this study, which reveal a ``stick'' feature extending out to its terminus at a broad arc to the left of the main galaxy. On the opposite side of the disk (right side of the upper panel in Figure~\ref{n4651}), additional shell-like\index{shells} features are clearly seen. \citet{foster2014} estimated the total stellar mass in the tidal debris to be $\sim4 \times 10^8 M_{\odot}$, constituting about 1/50th the stellar mass of NGC~4651. In addition, \citet{foster2014} obtained spectra of candidate globular clusters\index{globular clusters}, planetary nebulae\index{planetary nebulae (PNe)}, and HII regions\index{HII regions} that are spatially coincident with the substructures, and distinguished the kinematical signature of the accreted debris (including a possible progenitor core)\index{progenitors of streams} from the underlying galactic disk motions. The orbit\index{orbit fitting} derived from these is rather radial (as expected for an umbrella-like\index{umbrellas} remnant; see Chapter~6 of this volume), with pericenter of only a few kpc, apocenter of $\sim40$~kpc, and period of $\sim350$~Myr. This implies that the ratio of total mass of the progenitor and the whole of NGC~4651\index{NGC 4651} is $\sim0.15$, making this minor merger event analogous to the Sgr accretion in the Milky Way (see Chapter~2) and the Giant Stellar Stream\index{Giant Stellar Stream (GSS)} in M31 (Chapter~8). While detailed exploration of parameter space has yet to be achieved for this system, \citet{foster2014} did adapt an existing $N$-body model\index{N-body simulations} to show that some inferences can be made from analysis of the surface brightness and velocities\index{radial velocities} of the visible features and associated tracers.

As discussed in Chapter~7, the widths and surface brightnesses\index{surface brightness} of streams that are traced over a long portion of their orbits provide constraints on the number and sizes of dark matter subhalos\index{dark matter subhalos} in the host galaxy's halo \citep[see, e.g.][]{ibata2002,johnston2002,penarrubia2006,siegal_gaskins2008}. These authors show that the presence of dark matter subhalos in spiral galaxies would result in progressive heating of tidal streams as a result of close encounters, and in fact gaps\index{stream gaps} may be swept out of streams by interactions with subhalos (e.g., \citealt{yoon11, carlberg2013, erkal2014, ngan2014}). As \citet{penarrubia2006} pointed out, the average number of dark matter substructures (and, thus, the likelihood of encounters) in a Milky Way-like galaxy decreases monotonically from z $\sim$ 2 to the present, implying that ``old'' stream pieces like the ones visible in external galaxies are more likely to reveal perturbations than recently stripped ones. Tidal debris structures can also be used as kinematical tracers of the underlying gravitational potential in which they are produced \citep[e.g.,][; see Chapter~7 for detailed discussion]{johnston2001}. This has been attempted in the Milky Way using the Sgr streams\index{Sagittarius (Sgr) stream} (e.g., \citealt{law2005,penarrubia2005,law2010a}; see Chapter~2 of this volume for more discussion of Sagittarius). The use of streams beyond the Local Group may ultimately become fruitful for this purpose, as they can be traced over multiple wraps, providing much stronger constraints on the level of stream precession\index{precession of orbits} due to the flattening of the halo. Finally, old stream pieces stretching over multiple orbital wraps allows us to study metallicity gradients\index{metallicity gradient} in the stream (and thus within the progenitor galaxy)\index{progenitors of streams}, as has been done for the Sgr tidal stream \citep[e.g.,][Chapter~2 of this volume]{bellazzini2006, chou2007}. All of these techniques, and likely many others, applied to external galaxies provide valuable constraints on the hierarchical process of galaxy growth and evolution beyond what can be gleaned from our embedded perspective in the Milky Way.

\section{Stellar tidal streams as a galactic formation diagnostic}
\label{gx_formation_section}

One of the main objectives of stream surveys in nearby spiral galaxies is to compare the observations with cosmological simulations\index{cosmological simulations} to ascertain whether the frequency and surface brightness\index{surface brightness} of the detected stellar streams are consistent with those predicted by models.
Observational modeling and theoretical understanding of such diffuse and intricate features requires specifically tailored cosmological numerical simulations. 
There are two main difficulties for these models: 
 i) sufficiently fine mass and spatial resolution is needed to recover complex and delicate tidal features around Milky Way mass halos;
 and ii) a sufficient volume is required to build a statistically meaningful sample of host galaxies. 
To simultaneously meet these two requirements, state-of-the-art cosmological simulations are needed. 
For this reason, there are still a limited number of models of the stellar halos of Milky Way-like galaxies. 
Examples include the models by \citet[][also described in Chapter 5 of this volume]{bj2005} 
 and high resolution models of individual stellar halos for Milky Way-like galaxies based on the (dark matter only) Aquarius suite of simulations. 

Numerical simulations \citep[e.g.,][]{johnston08, cooper2010} can be used as a guide to what we may expect; models suggest that remnants of recent (0-8 Gyr ago) accretion events, 
 which correspond to the last few tens of percent of mass accretion for a Milky Way-like spiral, 
 should remain visible as substructures in presently observable stellar halos. 
These models also make predictions about what can be physically inferred from an external, ``global'' view of stellar halos. 
For example, from the shapes of debris remnants, we can infer the basic dynamical properties of the progenitor\index{progenitors of streams} -- \citet{johnston08} 
 demonstrates that tidal debris of different morphologies\index{morphologies of debris structures} each occupy different regions in the 
 time of accretion vs. orbital eccentricity/energy plane (see their Fig.~3), 
 and that surface brightness\index{surface brightness} also gives evidence of both the time of accretion 
 and the luminosity of a remnant's progenitor (see Fig.~4 of Johnston et al.).  
While this gives some general insight into properties of the accretion events, 
 there is considerable degeneracy in the inference of such properties, 
 which may best be thought of as providing reasonable estimates upon which to base specific modeling of the accretion events.

\begin{figure}[t]
  \begin{center}
   \includegraphics[height=1.0\textwidth, angle=270]{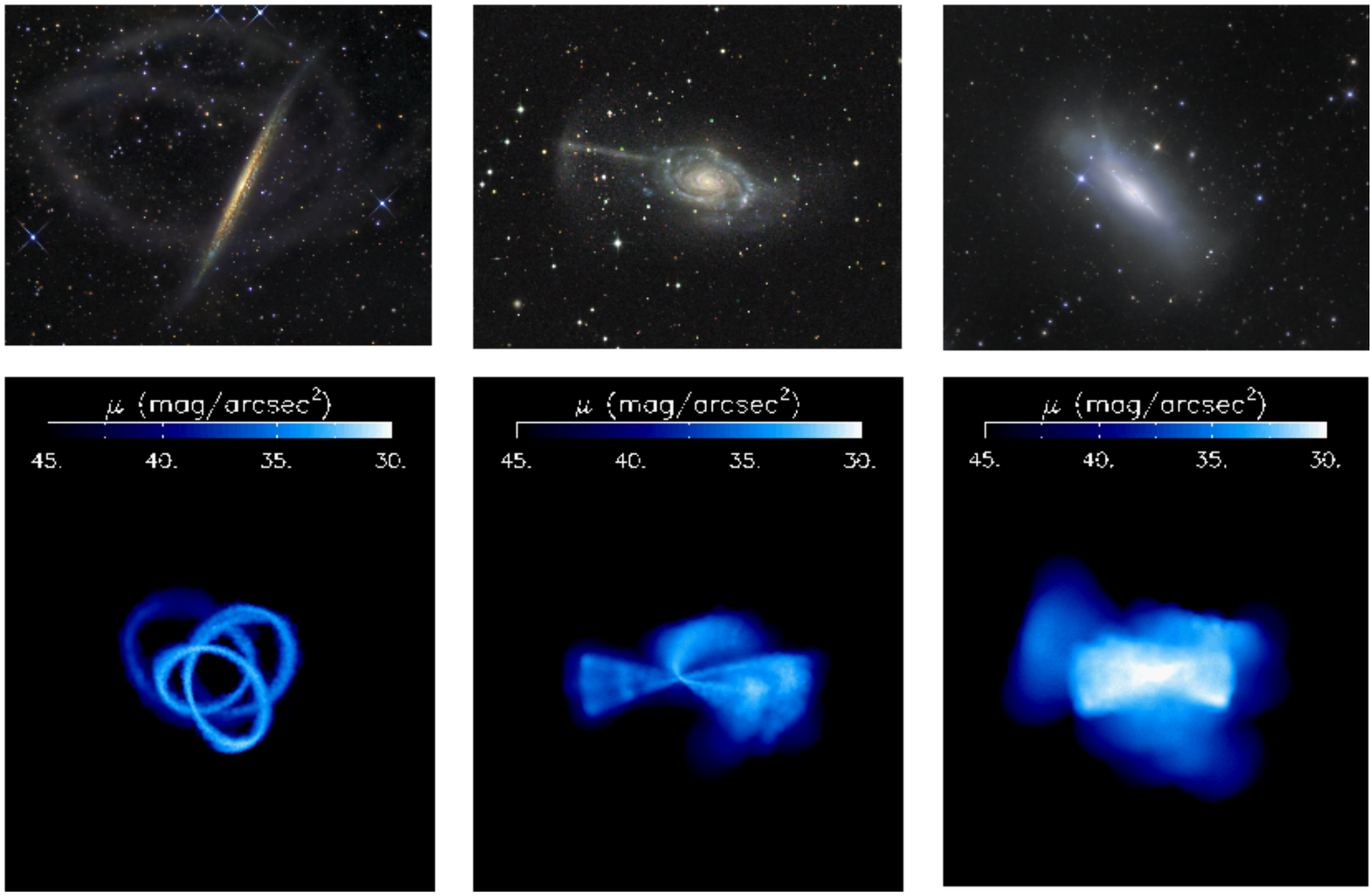}
    \caption{Bottom row: externally-viewed snapshots showing the surface brightness\index{surface brightness} of individual accretion relics in models of Milky Way-like stellar halos by \citet{johnston08}. The three main morphological\index{morphologies of debris structures} types identified in this study are illustrated: 
``great circle''\index{great circle streams} streams (or ``arcs''; left panel) arise from satellites accreted $\sim$6-10~Gyr ago on nearly circular orbits; ``cloudy''\index{clouds} morphologies (also dubbed ``shells''\index{shells} or ``plumes''; middle panel) arise from accretion events within the past $\sim8$~Gyr that fell in on eccentric orbits; and ``mixed''-type tidal remnants (right) arise from ancient  (more than 10 Gyr ago) accretion events that have had ample time to fully mix along their orbits. Top row: observational archetypes of each type of tidal debris from the survey by \citet{MD2010}: great circle stream in NGC~5907\index{NGC 5907} (left), shell-like features around NGC~4651\index{NGC 4651} (middle); and ``mixed'' debris near NGC~5866\index{NGC 5866} (right panel).}
    \label{johnston_sim_comp}
  \end{center}
\end{figure}

Figure~\ref{johnston_sim_comp} compares the predicted morphologies\index{morphologies of debris structures} of model debris structures from \citet{johnston08} to observed structures in external galaxies, 
 demonstrating that we see examples of the remnants predicted by the models in the local Universe. 
The classifications of debris structures suggested by Johnston et al. include: 
 ``great circles''\index{great circle streams} -- streams that arise from satellites on nearly circular orbits that were accreted $\sim6-10$~Gyr ago; 
 ``cloudy''\index{clouds} morphologies (also known as ``shells''\index{shells} or ``plumes'') resulting from recently-accreted (less than $\sim8$~Gyr ago) 
    satellites that were on rather destructive, radial orbits; 
 and ``mixed''-type tidal remnants from ancient accretion events (more than $\sim10$~Gyr ago) that have had time to phase mix\index{phase mixing} and become nondescript.
While these simulations make detailed predictions about the total number, frequencies, and specific properties of halo substructures, 
 there is no analogous observational data set to which these simulations can be compared en masse.
As demonstrated in Figure~\ref{johnston_sim_comp}, some observational examples of different stream morphologies\index{morphologies of debris structures} have been identified in the local Universe, 
 including ``great circles''\index{great circle streams} in NGC~5907\index{NGC 5907} \citep{shang98,MD2008} and M~63\index{M 63} \citep[see also Figure~\ref{n4013_m63}]{chonis11}, the ``plume''-like feature (or ``umbrella''\index{umbrellas}) in NGC~4651\index{NGC 4651} \citep{MD2010}, and a feature of ``mixed'' morphology around NGC~1055\index{NGC 1055} \citep{MD2010}. 

Generally, it is easiest to infer the physical properties of great circle streams\index{great circle streams}, 
 as the great circle is a reasonable tracer of the progenitor's orbit\index{orbit fitting}.  
Analytical relationships derived by \citet{johnston2001} may be used to estimate the accretion time 
 and total (dark matter+stellar+gas) initial mass of the progenitor\index{progenitors of streams} of a stream on a great circle. 
Surface photometry across streams in multiple filters can be used to infer the stellar populations 
 and total stellar mass for the stream (similarly to integrated light mapping of stellar mass surface density in external galaxies; 
 e.g., \citealt{zibetti09}). 
Variations in the optical colors along the stream can be used to infer changes in the mean properties of the stellar populations, 
 but are on the whole less diagnostic than similar studies using resolved stars in the Local Group. 
However, extraction of meaningful properties from the integrated light of streams requires S/N$>$5-10 above the local background fluctuations over a sufficiently wide area to cover the full stream. 
Thus, not only must the image be deep, but the backgrounds need to be well characterized. 
This can be seen in Figure~\ref{n4013_m63}, which shows that while hints of debris may be identifiable in SDSS\index{Sloan Digital Sky Survey (SDSS)} imaging 
 (e.g., \citealt{beaton2014}), extraction of physical properties requires deep, well-characterized imaging. 

\citet{cooper2010} coupled the Aquarius simulations to a state-of-the-art semi-analytic model known as GALFORM, which computes the properties (mass, size, star formation history and chemical abundance) of the galaxy forming in each dark matter halo. 
The GALFORM model is constrained through statistical comparisons to collective properties of the cosmological galaxy population (for example, optical and infrared luminosity functions), 
and by requiring that the surviving counterparts reproduce the observed size-luminosity relationship for Milky Way dwarfs\index{dwarf galaxies}. 
To meet these constraints, this technique demands fine-grained simulations such as Aquarius in order to adequately resolve the star-forming cores of satellite halos. 
This approach results in a set of dynamically self-consistent $N$-body realizations\index{N-body simulations} of stellar halos and their associated tidal streams at a resolution
 beyond the reach of current hydro-dynamical simulations \citep[e.g.,][]{abadi2006}, 
 and without the need to invoke many of the approximations required by previous models \citep[e.g.,][]{bj2005}. 
The individual star formation history of each satellite (and hence properties such as stellar mass, 
 luminosity and net metallicity) can be studied alongside the full phase-space evolution of its stars.

\begin{figure}[t]
  \begin{center}
   \includegraphics[height=1.0\textwidth, angle=270]{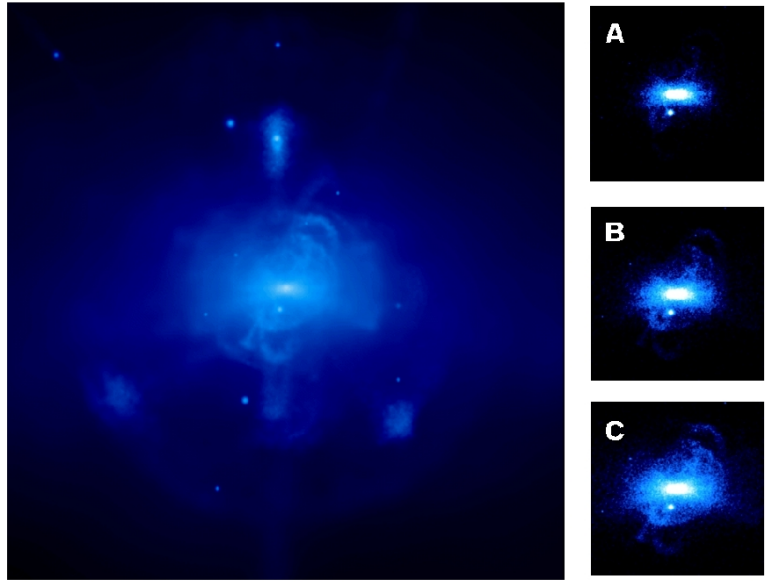}
    \caption{Expected halo streams around a Milky Way-like galaxy from a simulation \citep{bj2005}. 
 The figure shows an external perspective of one realization of a simulation within the hierarchical framework, with streams resulting from tidally disrupted satellites.
The snapshot on the left is 300~kpc on a side (the virial radius for a Milky Way sized galaxy), 
 and illustrates the result of a typical accretion history for a Milky Way-like galaxy. 
Right panels: theoretical predictions for the detectable tidal features in the same halo as the left panel, 
 but assuming three different surface brightness\index{surface brightness} (SB) detection limits: {\it A}: $\mu_{lim}$=28, {\it B}:$\mu_{lim}$=29 and {\it C}: $\mu_{lim}$=~30~mag~arcsec$^{-2}$. 
Each snapshot is 100~kpc on a side.
No discernible substructure is predicted for surveys with SB limits brighter than $\sim$27--28 mag~arcsec$^{-2}$ (e.g., POSS-II and SDSS\index{Sloan Digital Sky Survey (SDSS)}). 
This result also shows as the number of tidal features visible on the outskirts of spirals depends dramatically on the SB limit of the observations.
Moreover, the brightest substructures tend to be from the most massive satellites, which sample relatively rare accretion events \citep{johnston08,gilbert2009}.}
    \label{johnston_sim}
  \end{center}
\end{figure}

Current models have sufficient numbers of particles to resolve the main contributors of
 bright, coherent substructures that are similar to tidal features detected in the current imaging surveys \citep[e.g.,][]{MD2010}.
Thus it is possible to make sky-projected snapshots of these stellar halos from different viewing angles and a selected photometric band.
Each model halo has experienced a unique merging history and provides predicted surface brightness\index{surface brightness}, 
 morphologies\index{morphologies of debris structures} and overall distribution of the observable streams, survival of the 
 progenitors and stellar populations (or mean colors), that can be compared with observational data. 
These snapshots can be used as the input source for creating a mock catalogue of synthetic images, 
 which can be generated by adding simulated streams to real images including all the observational effects of the telescopes (e.g., typical sky noise, flat-field corrections, 
 surface brightness\index{surface brightness} limits, etc.) and contamination from other galactic substructures (e.g., the stellar disk that is not directly simulated by models of this type). 
An example of this technique using the \citet{bj2005} models as input is shown in Figure~\ref{johnston_sim}. 
These mock observations give preliminary predictions for the level of substructure detectable in the stellar halo of a nearby 
 spiral for the typical surface brightness\index{surface brightness} limits of current imaging surveys. 
Moreover, additional observational properties, like chemical compositions (both [Fe/H], [$\alpha$/Fe])\index{chemical abundances},
 can be explored as they relate to the other properties of substructure,
 the most significant being the time since accretion and luminosity (or stellar mass), as was done by, e.g., \citet{font2006}.

The \citet{bj2005} halos all have some structure visible at surface brightnesses\index{surface brightness} of $\sim$27-28~mag~arcsec$^{-2}$.
To surface brightnesses of $\sim$29~mag~arcsec$^{-2}$ there is $\sim$1 visible stream in each simulated halo, 
 and typically about 2 visible streams above $\sim$30~mag~arcsec$^{-2}$. 
The majority of the substructure, and thus the majority of the accretion history, is at surface brightnesses fainter than $\sim$30~mag~arcsec$^{-2}$. 
The degeneracy between the luminosity of the satellite, its accretion time, and its surface brightness\index{surface brightness} is studied in depth by \citet{johnston08}.
Our inability to see the fainter streams, which were either accreted earlier or come from lower luminosity progenitors\index{progenitors of streams}, 
 implies that our view of halo substructure beyond the Local Group will be dominated by the most recent and/or most massive accretion event,
 and, in either case, the most metal rich populations \citep{johnston08, gilbert2009}.
Moreover, more massive accretion events tend to preferentially populate the innermost regions of the halo ($R_{\rm proj} < 30$~kpc),
 as dynamical friction\index{dynamical friction}, which is more effective for more massive satellites, will cause the orbit of the progenitor to degrade \citep{johnston08}.
Thus, our view of extragalactic tidal streams at relatively shallow surface brightness limits (i.e., those of POSS and SDSS\index{Sloan Digital Sky Survey (SDSS)}) is highly biased
 to a specific subset of accretion events that are relatively rare for Milky Way sized galaxies.

\section{The role of interactions within dwarf galaxy halos}

Hierarchical formation models\index{hierarchical galaxy formation} predict substructure should exist on all scales, 
 not just around relatively massive galaxies like the Milky Way. 
Thus, exploration of the extended stellar structures of dwarf galaxies\index{dwarf galaxies}
 should reveal similar tidal features to those previously discussed. 
We first discuss the observational evidence for halos
 around dwarf galaxies and, in particular, for substructure in dwarf galaxy halos. 
Second, we discuss the impact that the creation of such substructures
 would have on dwarf galaxies. Understanding the formation and evolution of dwarf galaxies is particularly vital to form realistic ``initial conditions'' for simulations and to make highly detailed predictions of substructure properties. For many reasons, extra-galactic systems are best suited for these explorations.

\subsection{Observational evidence for substructure at dwarf galaxy scales} 

In the process of hierarchical structure formation\index{hierarchical galaxy formation}, it is likely that some of the most massive satellite dwarf galaxies\index{dwarf galaxies} themselves host their own, even smaller satellites\index{dwarf galaxies!satellites of} (dwarf spheroidals --``dSphs'' -- or globular clusters\index{globular clusters}). 
In the Milky Way, we know that the most massive classical dwarf galaxies have globular clusters associated with them -- specifically the Large Magellanic Cloud\index{Large Magellanic Cloud (LMC)}, the Small Magellanic Cloud\index{Small Magellanic Cloud (SMC)}, and the Fornax dSph\index{Fornax dwarf spheroidal (dSph)} \citep[e.g.,][]{forbes2000}, as well as the Sagittarius tidal stream\index{Sagittarius (Sgr) stream} \citep[e.g.,][]{law2010b}. 
The most massive Milky Way satellite, the Large Magellanic Cloud (LMC)\index{Large Magellanic Cloud (LMC)}, is itself part of a bound pair of dwarf irregular galaxies (with the Small Magellanic Cloud, or SMC\index{Small Magellanic Cloud (SMC)}). 
Thus, we would expect to find evidence of tidal interaction around dwarf galaxies\index{dwarf galaxies}, analogous to the features we see in halos of more massive galaxies. 
The majority of the dwarf galaxies in the Local Group are satellites, which have experienced interactions with their more massive host. 
Thus, it can be difficult to disentangle effects on the dwarf created during the accretion by its parent from those it experienced before falling in. 
Studying external, isolated dwarf galaxies may prove a more effective means of understanding the role of hierarchical formation,
 including the role interactions may have in forming the Hubble sequence at low masses.

In fact, many of the unique features of the LMC-SMC\index{Large Magellanic Cloud (LMC)} pair\index{Small Magellanic Cloud (SMC)} can be explained by their binary interaction \citep[e.g.,][]{besla2012},
 including the spectacular 200$^{\circ}$ Magellanic Stream\index{Magellanic Stream} in HI \citep{nidever2010}.
Binary pairs are somewhat rare, as \citet{robotham2012} find only two MW+LMC+SMC analog systems among the Galaxy Mass Assembly (GAMA) galaxies.
While only 3.4\% percent of GAMA galaxies are MW+LMC+SMC analogs,
 12\% of SDSS\index{Sloan Digital Sky Survey (SDSS)} galaxies have LMC-like companions \citep[i.e., luminous satellite within 75 kpc;][]{tollerud2011},
  which suggests that about one in four LMC-like satellites have a smaller SMC-like companion. 
The relatively low fraction of LMC-SMC binary satellite systems supports a ``transient'' nature,
  as detailed numerical simulations of the LMC-SMC suggest they may not remain a bound pair for long \citep{besla2012}.
In fact, it has been suggested (e.g., \citealt{donghia2008}) that satellites should often fall in as pairs or in groups\index{group infall}, rather than individually, 
 and the dwarf galaxies\index{dwarf galaxies} found at the edge of the Local Group (representing future accretions) are grouped \citep[e.g.,][]{mateo1998,mcconnachie2012}. 
Thus, we would expect to find evidence of tidal interaction around dwarf galaxies, analogous to the features we see in halos of more massive galaxies. 

Wide-field optical and near-infrared imaging has revealed stellar halos around many star-forming dwarf galaxies\index{dwarf galaxies} in and beyond the Local Group (see \citealt{stinson2009} for an extensive listing of many of these). 
While Stinson et al. determined that these extended stellar envelopes are not likely to arise due to tidal interactions with the (larger) host galaxies, it has not been determined whether they are the result of interactions with smaller satellites. 
Large area surveys of the most massive dwarf galaxies in the Local Group also indicate the presence of extended, ``halo-like''
 stellar populations at large effective radii, including M\,33\index{M33} \citep{2009Natur.461...66M}, the LMC\index{Large Magellanic Cloud (LMC)} \citep{nidever2007}, and the SMC\index{Small Magellanic Cloud (SMC)} \citep{nidever2011}. 
Further characterization, including full kinematic profiles, is necessary to determine if these extended structures are lower mass versions 
 of the halos found around Milky Way sized galaxies.

\begin{figure}[t]
  \begin{center}
   \includegraphics[height=0.9\textwidth, angle=270]{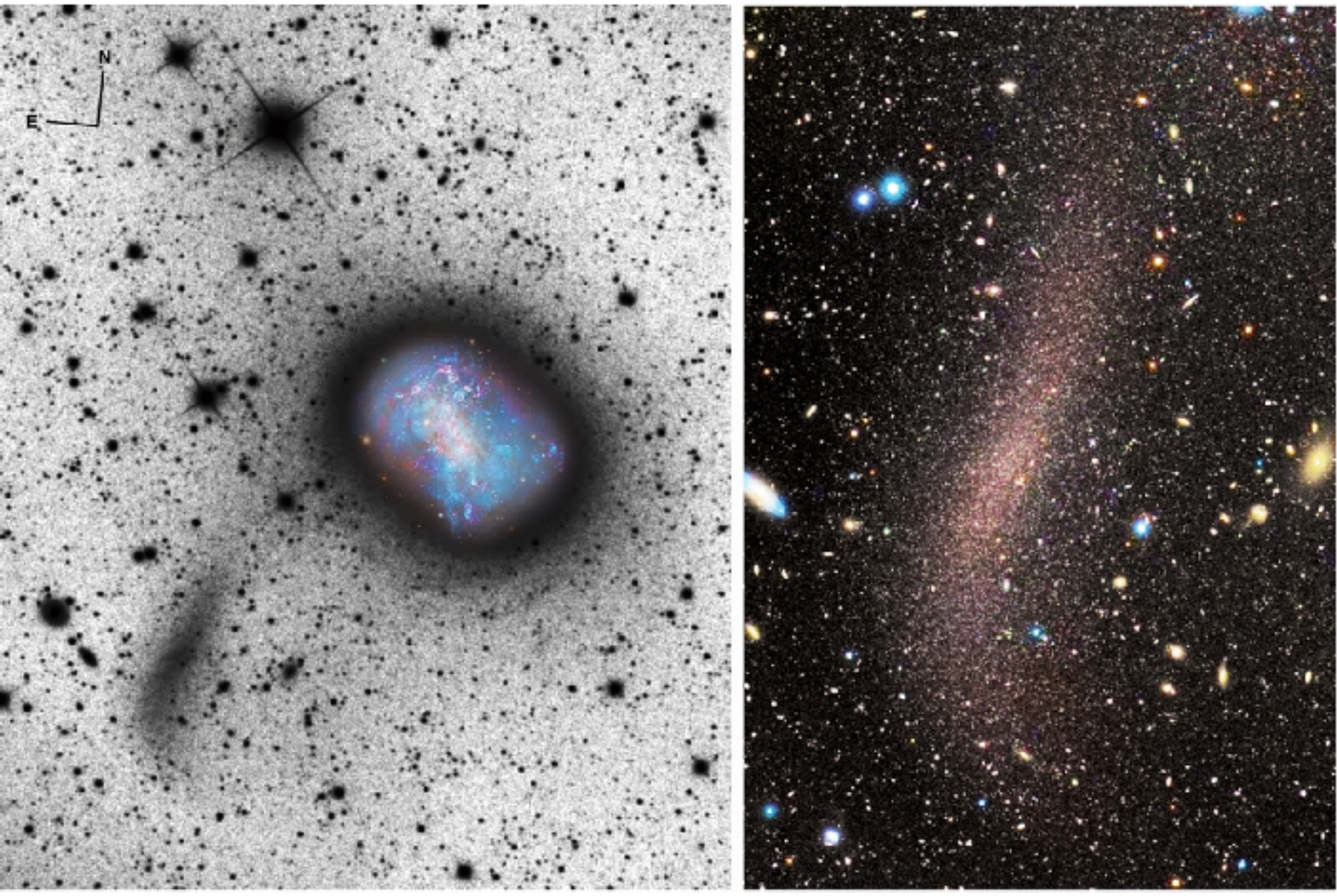}
    \caption{The stellar stream around the dwarf galaxy\index{dwarf galaxies} NGC~4449\index{NGC 4449}. 
 Left: greyscale image from \citet{MD2012} with a color inset for the galaxy, 
  which clearly indicates the presence of an $S$-shaped tidal stream approximately 7~kpc in length. 
  There is no clear association with any of the existing HI gas features \citep{hunter1998}. 
 Right: Subaru telescope sub-arcsec resolution image of the stellar stream \citep{MD2012}. 
  This is one of the few extra-galactic stellar streams that has been resolved into individual stars, 
   which provides direct probes of its stellar populations.}
    \label{n4449}
  \end{center}
\end{figure}

One of the many probable dwarf galaxies\index{dwarf galaxies} discovered by \citet{karachentsev2007} was an elongated feature near the 
 dwarf irregular galaxy NGC~4449\index{NGC 4449} in Digitized Sky Survey (POSS-II) plates (denoted as object ``d1228+4358'' in their catalog). 
NGC~4449 is a dwarf starburst galaxy with an irregular morphology, with luminosity ($M_{\rm V} = 18.6$) similar to that of the LMC\index{Large Magellanic Cloud (LMC)}, 
 but with much stronger and more widespread ongoing star-formation activity. 
Its cold gas and HII regions\index{HII regions} exhibit peculiar kinematics \citep{hartmann1986,hunter1998}, suggesting that it may have recently interacted with another galaxy. 
Using deep, wide-field imaging around NGC~4449\index{NGC 4449}, \citet{MD2012} definitively identified the Karachentsev et al. feature as a dwarf galaxy\index{dwarf galaxies!satellites of} undergoing accretion by NGC~4449 (see Figure~\ref{n4449}).
This new dwarf galaxy was also seen by \citet{rich2012} in a similar deep-imaging survey, which revealed the dwarf (dubbed NGC~4449B) and its $S$-shaped morphology that is characteristic of disrupting satellites. After fitting and subtracting a halo model, Rich et al. showed additional arcs and possible disk ripple features in the residual stellar surface brightness\index{surface brightness} maps, along with evidence for a break in the surface brightness profile of the NGC~4449 stellar halo. The morphology, size, luminosity, and surface brightness profile of the newly discovered stream/dwarf, along with evidence of tidal features in the NGC~4449 halo and outer disk, was suggested by \citet{rich2012} to result from a dwarf (NGC~4449B) that is on its first passage, and passed near the center of its host $\sim10^8$~yr ago. Thus, NGC~4449 is the first direct evidence of hierarchical structure formation similar to that seen in Milky Way-type galaxies, but on the mass scale of dwarf galaxies.

\subsection{Implications of dwarf-dwarf interactions}\index{dwarf galaxies!interactions between}

It is worth noting that the first stream from a dwarf-galaxy\index{dwarf galaxies} accretion event was found around one of the most intensely star-forming nearby galaxies. 
This leads one to wonder whether such accretion events are common among dwarf galaxies in recent epochs. 
It is possible that exact analogs to this stream have not been noticed in POSS or SDSS\index{Sloan Digital Sky Survey (SDSS)} images because they are uncommon. 
However, another explanation is that the majority of such structures are fainter, more diffuse, or at a larger radius than the NGC~4449\index{NGC 4449} stream, 
 and thus await future detection. 
If streams as in NGC~4449 are common around dwarfs, they re-ignite classic ideas about galaxy interactions triggering starbursts. 
Given the high rates of star formation in dwarf galaxies, it is natural to ask if satellites are responsible. 
Surveys along these lines have produced mixed results \citep{noeske2001, brosch2004, li2008}, 
but these studies were
 not looking for objects like the detected dwarf satellite of NGC\,4449 -- a gas poor, low-surface brightness analog to the Local Group dSphs -- and did not probe appropriate depths to find these objects.
Regardless of the implications for starbursts, evidence from NGC~4449 and the Fornax dSph\index{Fornax dwarf spheroidal (dSph)}, which shows traces of having swallowed a smaller dSph \citep{coleman2005}, suggests that accretion of even smaller building blocks is a viable avenue for direct assembly of dwarf galaxy stellar halos.

It has been proposed that dSphs orbiting massive galaxies such as the Milky Way may be the result of ``pre-processing''
 which is a term for the effects of interactions within groups of dwarf galaxies\index{dwarf galaxies!pre-processing in} \citep[e.g.,][]{donghia2009}.
More specifically, it has been suggested that the dSphs were once gas-rich, rotationally supported objects like the field galaxies (i.e., similar to NGC\,4449)
 whose properties were modified into the gas-free, dispersion supported dSphs via interactions with companions.
This is a compelling scenario for the class of dwarf galaxies in the Local Group known 
 as ``dwarf Transition'' objects \citep[see ][ among others]{mateo1998,grebel1999},
 whose properties are intermediate between those of dIrrs and dSphs.
Evidence of a dwarf-dwarf\index{dwarf galaxies!interactions between} interaction in NGC~4449\index{NGC 4449}, in conjunction with the known HI streams around this dwarf galaxy, may thus demonstrate such a process in action.
Moreover, \citet{nidever2013} identified an HI filament associated with the M\,31 dwarf satellite IC\,10\index{IC 10}. 
 The dynamics and orientation of the stream are inconsistent with the orbital parameters of IC\,10\index{IC 10}, 
 and \citet{nidever2013} suggest that the stream and other atypical HI features in the IC\,10 disk could be explained
 via interaction with a ``stealth'' companion.

\citet{wetzel2015} used the ELVIS simulations to explore the frequency of ``pre-processing'' for satellites within a simulated Milky Way host and found
  that nearly half of all satellites with stellar masses less than $10^{6}$ solar masses were pre-processed in a more massive satellite halo.
More generally, satellites with lower stellar masses or those closer to their host are more likely to have undergone pre-processing.
Recent observational work for the Milky Way and M\,31\index{Andromeda (M31) galaxy} also provides hints of associated satellites or debris,
 including potential satellites with similar line-of-sight velocities \citep[e.g.,][]{chapman2007,martin2009,tollerud2012}
 and potential kinematic associations between substructures \citep[e.g.,][]{deason2014}.
While these initial probes are tantalizing, full phase space realizations of these objects that are only permitted via precision distances and proper motions
 are required to fully explore these associations locally.
Moreover, though it is tempting to explain morphological transitions with dwarf-dwarf\index{dwarf galaxies!interactions between} interactions, 
 there is significant degeneracy with other physical processes that can alter HI morphologies even at large radius in a group environment,
 for instance the effects of ram pressure from the hot gaseous halo can be quite dramatic \citep[see case studies in][]{mcconnachie2007,kenney2014}
 and many isolated gas-poor dwarfs could be the result of ``fly-by'' interactions with their host \citep[see][]{teyssier2012}.
However, the HI debris created by dwarf-dwarf interactions and ram pressure are different,
 and finding more dwarf galaxies in ``distress'' will reveal the relative importance of these processes,
 which also have implications for their halo substructures.

A dramatic example of an ongoing dwarf-dwarf\index{dwarf galaxies!interactions between} interaction was seen by \citet{paudel2015}, who found 
 a pair of dwarf galaxies\index{dwarf galaxies} connected by a 15~kpc stellar bridge. 
The HI disk for one of the galaxies is ``completely destroyed'' and there are several knots of star formation 
 that have global properties similar to either young globular clusters or ultra-compact dwarf galaxies. 
The \citet{paudel2015} dwarf-dwarf merger bears a striking resemblance to a scaled down version of equal mass mergers 
 at larger total masses.
The importance of dwarf interactions in shaping stellar populations of low-mass galaxies is highlighted in a recent systematic, multi-wavelength study of the relative star formation rates in interacting pairs of dwarf galaxies (TiNy Titans, or TNTs) by \citet{stierwalt2014}. This work found clear evidence of star formation enhancement (by a factor of $\sim2.3\pm0.7$) among paired dwarfs relative to their unpaired counterparts. This enhancement occurs even in interacting pairs that are isolated by D~$>1.5$~Mpc from their nearest massive neighbor, showing that galaxy interactions are a frequent driver of enhanced star formation even outside the influence of larger galaxies. The Stierwalt et al. 2014 study also finds a factor of $\sim3$ increase in the fraction of paired dwarfs that are starbursting relative to single dwarfs, further highlighting the role of interactions in triggering star forming episodes. Thus far, though, stellar tidal debris signatures of these dwarf-dwarf\index{dwarf galaxies!interactions between} interactions have not been identified. [Note that this study is limited to pairs of mass ratio $(M_1/M_2)_* < 10$, while the disrupting companion to NGC~4449 has about 1/50th its stellar mass \citep{MD2012}.]

\begin{figure}[!t]
  \begin{center}
   \includegraphics[width=0.95\textwidth]{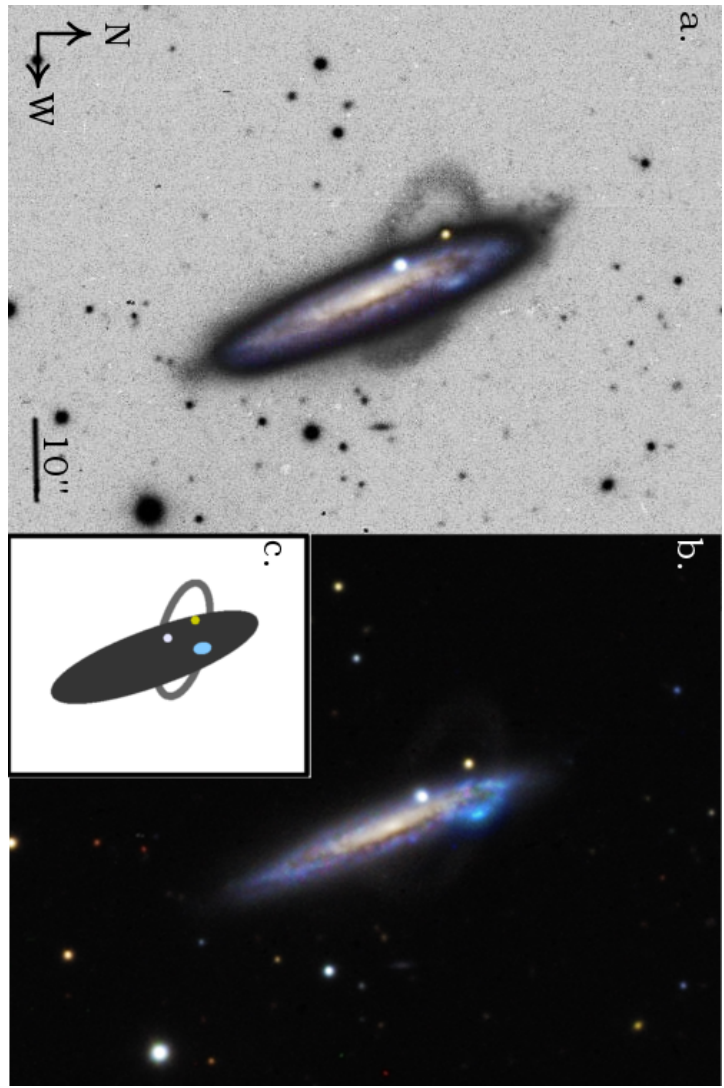}
    \caption{Images of the stream and associated blue overdensity in NGC~5387\index{NGC 5387}. The left panel is the $R$-band image from the VATT, with the SDSS\index{Sloan Digital Sky Survey (SDSS)} color image inlaid in the central regions, and clearly shows a full wrap of tidal debris about the disk of NGC~5387. The right panel shows a merged image composed of the SDSS, VATT, and $GALEX$ FUV images, which more clearly highlights the blue overdensity near the northernmost edge of the disk of NGC~5387. The inset in the right panel is a schematic highlighting the disk in dark gray, the stream as a lighter gray loop, the blue overdensity as a blue circle, and two foreground stars. [Reproduced from \citet{beaton2014}.]}
    \label{n5387}
  \end{center}
\end{figure}

\section{Induced star formation, disk structure, and tidal streams}

Simulations of structure formation on galactic scales in the prevailing $\Lambda$CDM paradigm predict that Milky Way-sized host galaxies 
 should accrete several massive satellites from z~$\simeq$~1 to the present. 
We now consider the effects of these interactions on the disk of the parent galaxy. 
Given the significantly shorter dynamical timescales in the disk,
 the potential effects of a even a minor accretion event could be profound.
In particular, such events could incite abnormal structures in dissipational HI disks,
  either creating features after a pass through
  or creating large scale, global distortions in the HI structure.
In the stellar component of the disk, the effects could be both more subtle
 and more long lived.

One example of a burst of star formation in a Milky Way-like galaxy with an associated recent accretion event is seen in NGC~5387\index{NGC 5387} \citep[note that NGC~5387 can be considered a Milky Way analog based on its properties, but contains about an order of magnitude less mass than the MW]{beaton2014}. This system shows a low surface-brightness\index{surface brightness} feature in SDSS\index{Sloan Digital Sky Survey (SDSS)} imaging, that was explored with deeper imaging by Beaton et al., who found a stellar stream extending over at least an entire orbital wrap, with median surface brightness of $\sim$24.5~mag~arcsec$^{-2}$ in $R$-band (see Figure~\ref{n5387}). This stream has a redder color than the typical stellar populations of NGC~5387, and contains a total stellar mass of $\sim6 \times 10^8 M_{\odot}$. Coincident with the position of the stream's crossing of the NGC~5387 disk is a ``blue overdensity'' that is not only blue in optical colors, but is producing a large FUV flux as well. Evidence from the FUV flux as well as follow-up spectra obtained by \citet{beaton2014} suggests a very recent ($\sim8$~Myr ago) star formation event of total stellar mass $2.5\pm1.3 \times 10^7 M_{\odot}$. Beaton et al. conclude that this blue overdensity is a complex of multiple HII regions\index{HII regions} produced by star formation either induced in the disk by the minor merger event, or in the dwarf galaxy\index{dwarf galaxies} progenitor\index{progenitors of streams} of the stream itself. Whichever scenario is shown to be true, it is clear that the blue overdensity and its associated tidal stream represent star formation induced by the accretion of a satellite about 1/50th the mass of its host. 

The presence of tidal streams encircling galaxies with warped disks (see, e.g., \citealt{MD2008}) may suggest satellite galaxy perturbations as the origin of those disk features (e.g., \citealt{velazquez_white1999,weinberg_blitz2006}). This suggests that promising galaxies to search for extragalactic tidal streams are those that display disk asymmetries in optical or HI images that may result from gravitational interaction with the tidally disrupting companions. One striking case of such a system is NGC~4013\index{NGC 4013}, an isolated spiral galaxy with a prominent HI warp \citep{bottema1987} that has been revealed by deep imaging \citep{MD2009} to contain a faint loop-like stellar tidal stream at fairly low inclination to the disk (see the upper panels of Figure~\ref{n4013_m63}).
The sky-projected morphology of this structure displays a remarkable resemblance to an edge-on view of models of the Monoceros Ring\index{Monoceros Ring} feature (see Chapter~3) in the Milky Way as a tidal debris structure \citep{penarrubia2005}. This suggests that the progenitor\index{progenitors of streams} system of the NGC~4013 stream may have been a low-mass satellite on a low-inclination, nearly-circular orbit that was accreted approximately $\sim2.8$~Gyr ago.
Stellar streams have also been discovered in the warped spiral galaxies NGC~5907\index{NGC 5907} \citep[see also Figure~\ref{n5907}]{MD2008} and M~63\index{M 63} \citep[see also Figure~\ref{n4013_m63}]{chonis11}, showing that disks that are apparently undisturbed as seen in the optical, but warped in HI maps, may reveal signatures of recent accretion events in deep imaging surveys

It has been shown that such accretion events should lead to strong warping, flaring, and thickening of an initially cold stellar disk \citep{kazantzidis2008}, 
 and to the generation of bars and spiral structure (e.g., \citealt{toomre1972, gauthier2006, dubinski2009, purcell2011}). 
Such perturbations may also lead to the formation of long-lived, ring like stellar features in the outer reaches of the disk 
 that may extend several kiloparsecs from the disk plane, and have surface brightnesses\index{surface brightness} in the range of 25-30 mag arcsec$^{-2}$ 
 (see Fig.~6 of \citealt{kazantzidis2008}). 
Indeed, observations in the Milky Way are uncovering wave like perturbations in the disk\index{Galactic disk} in stellar densities\index{Galactic disk!oscillation of midplane} \citep{widrow2012, yanny_gardner2013, xu2015} 
 and velocities\index{Galactic disk!velocity substructure} \citep{gomez2012a, gomez2012b, carlin2013, williams2013}, 
 with an accompanying array of simulations predicting the formation of such features 
 (e.g., \citealt{chakrabarti2009, micheldansac2011, purcell2011, gomez2013, faure2014, widrow2014}). 
Evidence for dynamically heated populations is not unique to the Milky Way -- \citet{dorman2013} have identified a kinematically cold population in the halo of M31\index{Andromeda (M31) galaxy}.

\section{Future prospects}

While the study of tidal streams from major mergers/encounters (1:3 mass ratio) or even minor encounters (1:3 to 1:10 mass ratio) is an old field, the extension of these studies to those streams formed by satellite galaxies, a.k.a., micro-mergers (mass ratio $<$~1:10), is a relatively new area of exploration.
Study of mergers on this mass scale provide a direct way of addressing some open questions on galactic formation and evolution. 
In the last decade, the observational effort has yielded an unprecedented sample of bright stellar streams in nearby spiral galaxies, 
 including the discovery of observational analogs to the canonical morphologies\index{morphologies of debris structures} found in $N$-body models\index{N-body simulations} of stellar halos (\citealt{johnston08}; see chapter 6). 
This offers a unique opportunity to study in detail the apparently still dramatic last stages of galaxy assembly in the local universe and to probe the anticipated estimates of frequency of tidal stellar features from the $\Lambda$-CDM paradigm for MW-sized galaxies. 
Moreover, these discoveries demonstrate the need for deep, wide-field imaging that pushes fainter than current surveys in order to visualize 
 external galaxy halos on par with the highly substructured portrait of our own Milky Way and M\,31\index{Andromeda (M31) galaxy}.
Such studies will address the following key questions (among others) on several aspects of hierarchical galaxy formation\index{hierarchical galaxy formation}:

\begin{itemize}

\item How many tidal debris features still exist as
recognizable substructures in nearby spiral galaxies?

\item Is the abundant number of
stellar streams exceptional in the Local Group spirals, and are tidal streams as common as predicted by cosmological models? 

\item At what rate are stellar streams still forming in the local Universe? 


\item Are the tidal stream properties (e.g., mean surface brightness) observed in the local Universe consistent with predictions from $\Lambda$-CDM simulations? 

\item What can we learn about baryonic processes within dark matter halos from observations of these stellar halo substructures? 

\item What is the (stellar) mass spectrum of the streams, and hence the mass spectrum of the progenitor satellites? 

\item What is the fraction of halo stars attributable (within data limitations) to distinct structures?

\item Do all dwarf galaxies contain evidence of hierarchical accretion, and what is the role of such interactions in forming the progenitors of streams in larger galaxies?

\item Do streams (or their progenitor satellites) contribute to disk heating and the formation of morphological perturbations observed in nearby galaxies?  

\item What is the incidence of low-inclination streams and what is their role in re-shaping the outer disks of nearby spiral galaxies?

\end{itemize}

\vspace{0.2cm}

The study of external tidal streams also has the potential to tackle a significant number of other topics that are the focus of current astrophysical research (e.g., stellar populations of halos, the resilience of the disks involved with minor mergers, accretion of globular clusters\index{globular clusters}, induced star formation in streams, near-field cosmology, satellite dynamics, dark matter halo shapes, etc.). In particular, the interpretation of global properties of galaxy halos and outer disks from 
resolved stellar populations (from, e.g., the Hubble Space Telescope\index{Hubble Space Telescope (HST)} survey GHOSTS; \citealt{radburn_smith2011}, or the CALIFA high-resolution spectroscopy survey of nearby stellar systems; \citealt{sanchez2013}) requires understanding the role and prevalence of tidal debris in galaxy halos.
In addition, studying stellar population gradients along tidal streams via deep HST photometric data (see, e.g., \citealt{aloisi2005}) will render important constraints on the effect of tides on the stellar formation history of dwarf galaxies\index{dwarf galaxies}. The panoramic view of tidal streams in external galaxies also offers an excellent opportunity to demonstrate tidal stripping of globular clusters formed in satellite galaxies, which may correspond to an important fraction of the globular cluster\index{globular clusters} population of the host, as earlier proposed by \citet{sz78}. Ultimately, the ideal scenario would require resolving stellar populations in large numbers of galaxies at distances of 10-20 Mpc, which will be feasible in the next one or two decades with thirty-meter class ground-based telescopes or the proposed suite of space-based instrumentation.

Finally, the future census of tidal streams and their properties will also provide an essential framework for exploring whether the Milky Way is a template for the archetypal spiral galaxy. The next generation of galactic surveys (LSST)\index{Large Synoptic Survey Telescope (LSST)} and future astrometric space missions (Gaia)\index{Gaia} will dissect the structure and formation of the Milky Way with unprecedented detail, leading to a revolutionary improvement of our understanding of the Galaxy.  In this regard, the study of these structures in external systems will be complementary in interpreting this local Galactic archaeological data in the context of galaxy formation and evolution, providing unique data in order to quantify how typical the Milky Way is with respect to other nearby galaxies of its type.

\begin{acknowledgement}
JLC gratefully acknowledges support from the NSF under grants AST 09-37523 and AST 14-09421.
\end{acknowledgement}

%
\end{document}